\documentclass[aps,pra,twocolumn,showpacs,superscriptaddress,groupedaddress]{revtex4-1}

\usepackage{amsmath,braket,amssymb,natbib,graphicx,float,amsthm,xcolor,outlines,mathrsfs,multirow,hhline,svg}

\usepackage{dcolumn}
\usepackage[english]{babel}
\usepackage[utf8]{inputenc}
\usepackage[T1]{fontenc}
\usepackage{listings}
\usepackage[colorinlistoftodos]{todonotes} 
\usepackage[colorlinks=true, allcolors=blue]{hyperref}
\usepackage{hyperref}
\usepackage{comment}
\begin{document}
\title{Nonlinear Stochastic and Quantum Motion from Coulomb Forces}
\author{Luca Ornigotti}
\email{luca.ornigotti@gmail.com}
\affiliation{Department of Optics, Palack\'{y} University, 17. listopadu 1192/12, 771 46 Olomouc, Czech Republic}
\author{Darren W. Moore}
\email{darren.moore@upol.cz}
\affiliation{Department of Optics, Palack\'{y} University, 17. listopadu 1192/12, 771 46 Olomouc, Czech Republic}
\author{Radim Filip}
\email{filip@optics.upol.cz}
\affiliation{Department of Optics, Palack\'{y} University, 17. listopadu 1192/12, 771 46 Olomouc, Czech Republic}

\begin{abstract}
 Controllable nonlinear quantum interactions are a much sought after target for modern quantum technologies. They are typically difficult and costly to engineer for bespoke purposes. However controllable nonlinearities may have always been in reach via the natural and fundamental forces between quantum particles. The Coulomb interaction between charged particles is the simplest example. We show that after eliminating the harmonic part of the Coulomb force by an auxiliary linear force, the remaining reciprocal nonlinear part results in a directly observable non-reciprocal nonlinear effect: increase of the signal-to-noise ratio (SNR) of the coherent displacement of one particle, driven by the position noise, or uncertainty in quantum regime, in another particle. This essential evidence of nonlinear forces is present across large ranges of trap frequency and mass scales, as well as visible in both stochastic and quantum regimes.
\end{abstract}

\maketitle

\section{Introduction}

The control of quantum systems has proceeded apace and many experimental settings possess precise control over linear quantum systems~\cite{weedbrook_gaussian_2012,adesso_continuous_2014}, as in quantum optics~\cite{cerf_quantum_2007,asavanant_optical_2022} and quantum optomechanics~\cite{noauthor_quantum_2015}, and the most simple nonlinear systems such as the Jaynes-Cummings model, exemplified in cavity-QED~\cite{blais_circuit_2021} or trapped ion systems~\cite{monroe_programmable_2021}. To construct a large scale nonlinear system from such simple nonlinear systems is a current challenge and therefore many experiments are pushing into exciting nonlinear regimes where standard modes of analysis fail, often catastrophically. Such nonlinear regimes are, as might be expected, also the site of yet uncovered quantum phenomena and absolutely required for the most advanced quantum technologies to come to fruition~\cite{calcluth_sufficient_2024,asavanant_multipartite_2024}. Similar statements can be made regarding recently developed platforms operating entirely in the classical regime and near the classical-quantum boundary, such as levitated nano-objects~\cite{Timberlake2019,Vinante2020,Conangla2020,Svak2021,Tebbenjohanns2021,Rieser2022,Hofer2023,Gutierrez2023,Brown2023,Bykov2023,Brzobohaty2023,reisenbauer_non-hermitian_2024}. Such systems are on the road from being inherently stochastic in a high temperature environment to gradually reaching semiclassical and quantum domains. The resulting classical-to-quantum transition with macroscopic objects will likely open access to unexplored physics once they enter nonlinear regimes, such as entanglement-by-heating~\cite{laha_entanglement_2024}, and with the possibility of applications in topics such as quantum sensing for parameter estimation~\cite{ivanov_quantum_2022,mahmoudi_optimal_2024} or of gravity~\cite{carlesso_testing_2019}, quantum thermodynamics~\cite{martinez_dynamics_2013,ray_thermodynamic_2023} and even quantum computing~\cite{moore_quantum_2019,vandre_graphical_2024}.

In fact the most prospective systems work in a regime derived from the natural forces between the particles. Perhaps the most widely exploited is the direct reciprocal interaction by the Coulomb force between a pair of charged particles. With state of the art technology such interactions can be controllably realised on large ranges of mass and frequency scales, from charged levitated nanoparticles~\cite{rudolph_force-gradient_2022,penny_sympathetic_2023,deplano_coulomb_2024} all the way down to trapped individual ions~\cite{nguyen_experimental_2021} and even recently between pairs of electrons~\cite{Beysengulov2024}. When confined by effective harmonic traps, the linearisation of the force results in coupled harmonic oscillations, with each directional mode decoupled from the others~\cite{deng_quantum_2008}. We promote an approach to nonlinear effects which show themselves in the intermodal coupling achievable by going beyond the harmonic approximation. Such effects have already been observed in the rotating wave approximation for trapped ion systems~\cite{ding_quantum_2017,ding_quantum_2018,maslennikov_quantum_2019}. However, the steps beyond this approximation in macroscopic mechanical systems have not been exploited yet. On the other hand, in the context of microscopic single ion heat engines cubic interactions between different radial and axial modes have been engineered by tailoring the trap geometry~\cite{abah_single-ion_2012,rosnagel_single-atom_2016}.

As the first step we investigate directly observable nonlinear effects arising from fundamental forces beyond the harmonic approximation and outside any rotating wave approximations, in both classical stochastic and quantum mechanical regimes. In essence, the nonlinear effects are made manifest non-reciprocally through the noise or uncertainty stimulated properties of the system, in our case a coherent displacement of one particle driven via the noise or uncertainty in the other. By definition, this can only occur via nonlinear interaction of the particles. Even in an approximation where the intermodal effects along different directions are suppressed, this effect can be predicted by expanding the Coulomb interaction only to third order, the first nontrivial nonlinear term, and focusing on a single nonlinear interaction along the straight line between the particles. The interaction is compound, maintaining the reciprocity of the Coulomb force, but does not prevent the observation of non-reciprocal effects in classical and quantum regimes. 

In what follows we will make these propositions more precise by introducing the simplest model, presenting testable results, then demonstrating how such effects can be understood in terms of a suitable approximation. The detection of nonlinear motion, quantified by a noise/uncertainty driven increase in the signal-to-noise ratio of the momentum displacement, demonstrates the activity of the nonlinear part of the Coulomb interaction. These effects are shown to be observable across a range of experimentally relevant parameters. This forms a proof-of-principle step towards experimental verification in the stochastic and, later, the quantum regime. We expect that the observation of such effects will open further directions for further investigations beyond Gaussian entanglement~\cite{rudolph_force-gradient_2022,penny_sympathetic_2023}.

\section{Results}

\subsection{Nonlinear Motion From the Coulomb Force}

\begin{figure*}[t!]
\centering
\includegraphics[width=\textwidth]{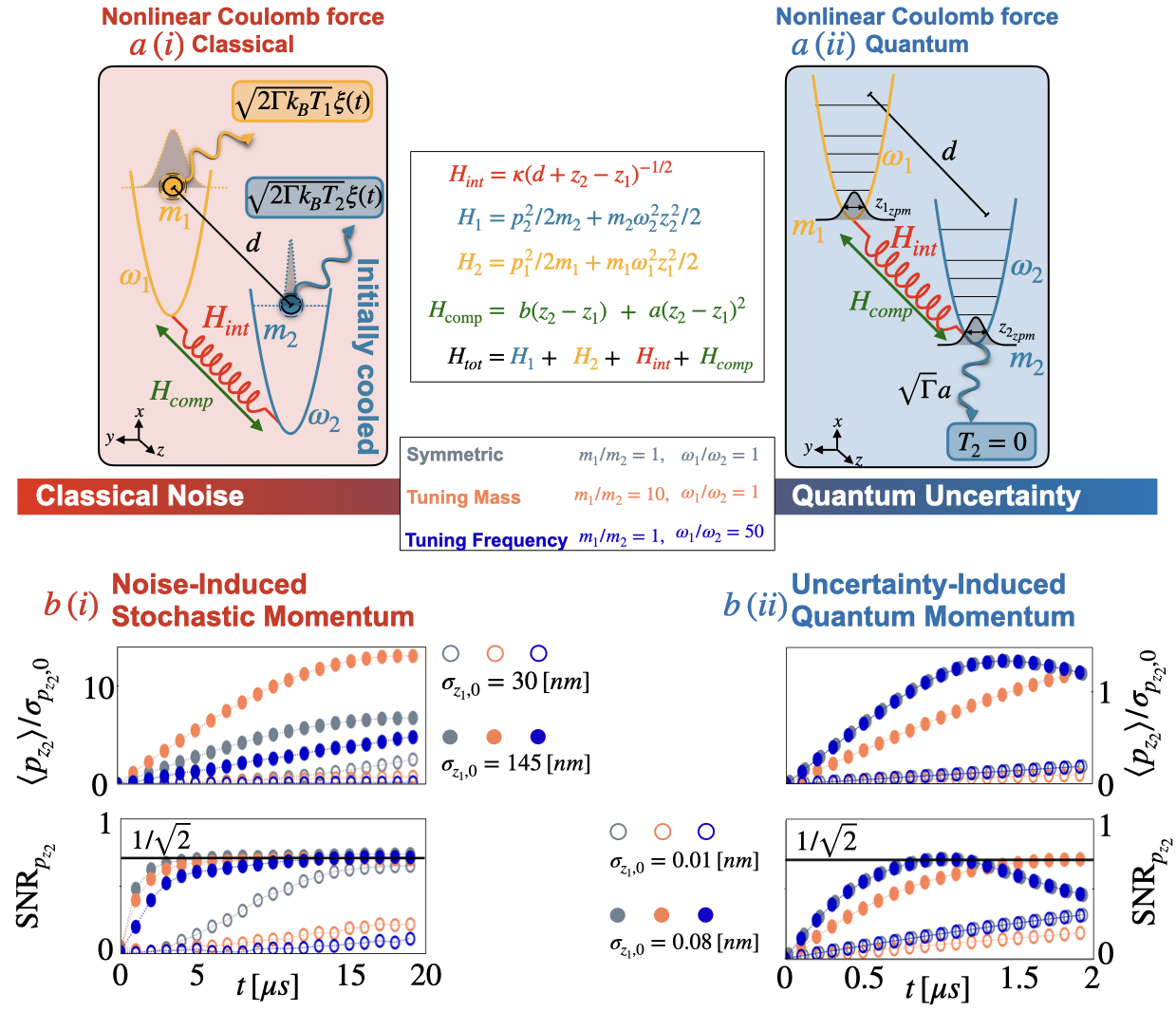}
\caption{\label{fig0}\small{
{\bf Noise and uncertainty-induced momentum displacement via a nonlinear Coulomb interaction.}
{\bf a (i)}, {\bf a (ii)} Illustration of the principle idea in classical and quantum regime. Two harmonically confined particles (blue, yellow) experience a nonlinear Coulomb interaction (red spring) along a single axis, together with a compensating force (green). In the classical regime, the stochastic particle $1$ is prepared in oscillator equilibrium states via dissipation to thermal environment at temperature $T_1$, while particle $2$ is initially cooled to $T_2'=10$~mK  in a room temperature environment $T_1=T_2 = 300$~K.
In the quantum regime, the particles are prepared in ground states and the only fluctuations arise from the quantum uncertainties. A weak linear damping with a rate $\Gamma$, acting only on particle $2$, is present in order to ensure the stability of the effect.
The trap frequency modulation, and mass disproportion of particle $1$, as showed in the table, are used to generate unidirectional flow of fluctuations to particle $2$, thus avoiding back-actions.
{\bf b (i)}, {\bf b (ii)} Time evolution of mean momentum $\langle p_{z_2} \rangle$, normalised to the initial standard deviation (top) and signal-to-noise ratio $\mathrm{SNR}_{p_{z_2}}$ (bottom).
For large noise (full circles), the $\mathrm{SNR}=1/\sqrt{2}$ is quickly reached by all regimes, but tuning frequency (blue) allows for better noise control. At lower noise (empty circles), the parametric symmetry is the only regime reaching the $\mathrm{SNR}$ bound (grey). In the quantum regime ({\bf b} (ii)), the ground state fluctuations (empty circles) are equally harnessed by all regimes, whereas an initial uncertainty amplification, by freefall, (full circle) allows the $\mathrm{SNR}$ bound to be reached by all regimes. Symmetric (grey) and Tuning Frequency (blue) further experience the faster uncertainty growth ($\mathrm{SNR}$ drop), not visible for tuning mass (orange), which is also the best regime here.}}
\end{figure*}

A pair of equally charged particles of masses $m_i$, $i\in\{1,2\}$, confined to a three dimensional harmonic trap and interacting via the Coulomb potential, can be arranged along the $z$ axis by having a much greater trap frequency in that direction~\cite{Rieser2022,Brzobohaty2023,reisenbauer_non-hermitian_2024}.
The axes can then be chosen so that the equilibrium points are $\mathbf{d}^{(i)}=\begin{pmatrix}0 & 0 & z_{i,0}\end{pmatrix}$. Even beyond the harmonic approximation, where the motional axes are not decoupled, the interactions depend upon the distances between the two particles along the respective axes. If motion along the $x$ and $y$ axes is sufficiently cooled so that fluctuations along these axes are small, then the interactions between $x$ or $y$, and $z$ are suppressed. The Hamiltonian of the system can then be written as
\begin{equation}
    H=\frac12\sum_i\left(\frac{p_i^2}{m_i}+m_i\omega_i^2(z_i-z_{i,0})^2\right)+\frac{\kappa}{\sqrt{(z_1-z_2-d)^2}}\,,
\end{equation}
where $\omega_i$ are the trap frequencies, $\kappa = q_1 q_2/4\pi \epsilon_0$ is the coupling strength of the reciprocal interaction calculated via the charge $q_i$ and the electric permittivity in vacuum $\epsilon_0$, and $d=z_{1,0}-z_{2,0}$ the initial distance between the particles' equilibrium positions. The setup is illustrated in Fig.~\ref{fig0}a. In the harmonic approximation, expanding around the distance between the particles in each direction to second order, results in an extra displacement term in the $z$ direction for both particles. The motion of the charged particles is then described by harmomically coupled oscillators with modified frequencies. The coupling induced between the particles is intramodal such that the modes along different axes do not talk to each other.
The third order term introduces nonlinear and intermodal interactions. To isolate them, a compensating force eliminating the lower order contributions is necessary. The constant force can be compensated via linear tilt through an electrostatic force~\cite{ciampini_experimental_2021}, whereas the linear force can be compensated by parametric control, such as feedback~\cite{Setter2018,Conangla2019,Magrini2021}.
In the regime of optimal compensation, assumed throughout the manuscript, the still reciprocal interaction Hamiltonian can be approximated as
\begin{equation}\label{eq2}
    H_3\approx\frac{\kappa}{d^4}(z_1-z_2)^3=\frac{\kappa}{d^4}\left(z_1^3-z_2^3+3(z_1z_2^2-z_1^2z_2)\right)\,.
\end{equation}
A non-optimal compensation still results in the non-reciprocal nonlinear effect, albeit reduced in visibility, see SM Note 3. Due to the nonlinearity of the Coulomb force a reciprocal cubic interaction emerges along the $z$ axis. This is a minimal model of the nonlinear Coulomb interaction between two particles. While similar nonlinear interactions can be constructed for ions directly via the trap geometry for hybrid radial and axial modes~\cite{abah_single-ion_2012,rosnagel_single-atom_2016}, here the nonlinearity emerges directly from the Coulombic interaction between the particles. Both the trap and the compensation operate in the harmonic approximation, and therefore cannot contribute to the nonlinear effects outlined below.
Differently than for optical cubic potentials~\cite{rakhubovsky_stroboscopic_2021}, the interaction in Eq.~\ref{eq2} combines competing cubic nonlinearities. That is, the cubic single particle potential $z_1^3,\,z_2^3$, and the cubic interparticle potential $z_1z_2^2,\, z_1^2z_2$.
Their vying nature may limit the direct observation of the interparticle nonlinear noise or uncertainty induced phenomena.
Quantisation can be accomplished by promoting the canonical variables to operators satisfying the commutation relations $[z_i,p_i]=i$. 

\subsubsection{Classical Noise-Induced Momentum}

\begin{figure*}[t!]
\centering
\includegraphics[width=\textwidth]{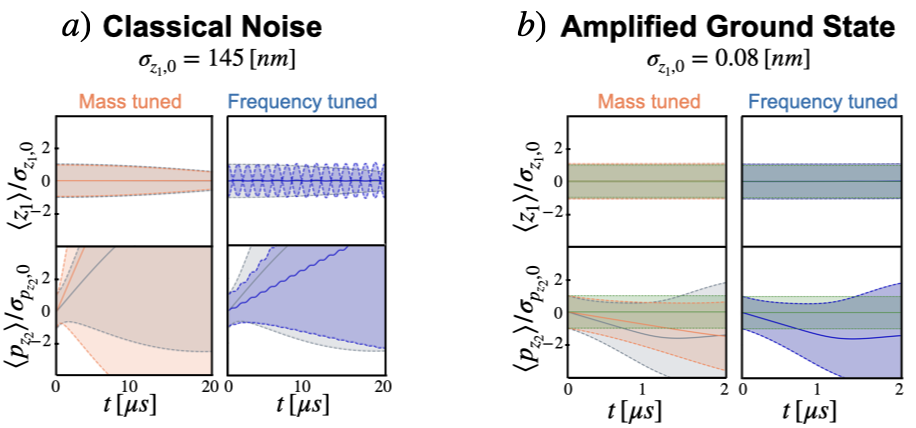}
\caption{\label{fig1p2}\small{
{\bf Analysis of time evolution of variables undergoing nonlinear motion.} Analysis of time evolution of position $z_1$ and momentum $p_{z_2}$ at different initial fluctuations of $z_1$. The shaded area represents the standard deviation around the mean evolution (solid). 
All quantities are normalised to the standard deviation of their respective initial states.
Symmetry breaking by mass tuning (orange) and frequency tuning (blue) allows to control divergence in $p_{z_2}$ in both mean and standard deviation. In the classical regime (a) the mass tuning visibly performs better than the other strategies as it produces larger momentum drift $\langle p_{z_2} \rangle$.
For the quantum regime (b), the symmetric (grey) and frequency tuned (blue) outperform the mass tuned with the same metric. It confirms the result presented in Fig.~\ref{fig0} ($b$ (i),$b$ (ii)).}}
\end{figure*}

Nonlinear interactions such as $z_1^2z_2$ allow for the possibility to coherently displace the momentum of one particle via increases in the initial position noise of the other. Heuristically, the complete reciprocal interaction term of Eq.~\eqref{eq2} indicates that for initially uncorrelated states the mean momentum displacement in one mode is rapidly driven by the noise in the other i.e. $\langle p_{z_2} \rangle \approx \langle p_{z_2,0} \rangle -3\kappa (\langle z_{1,0}^2 \rangle - 2\langle z_{2,0} \rangle \langle z_{1,0} \rangle + \langle z_{2,0}^2 \rangle)t/d^4$.
The reciprocal nature of the interaction Hamiltonian of Eq.~\eqref{eq2} suggests that a separate asymmetrical manipulation performed solely on one of the two particles can enhance the inter-particle non-reciprocal noise-induced nonlinear effect observed on the other. 
For parametrically symmetric interactions, $m_1=m_2$ and $\omega_1=\omega_2$, the minimal asymmetrical manipulation is accomplished by unbalancing the initial distribution of thermal noise. That is, prepare the initial thermal state at temperature $T=300$~K of particle $2$ at an effective temperature of $T_2' = 10$~mK by cooling, while particle $1$ is prepared in an oscillator thermal-equilibrium state at room temperature $T_1=300$~K.
These are zero-mean Gaussian states with variances in position $\sigma_z^2=k_BT/m\omega$ and momentum $\sigma_{p}^2=mk_BT$, where $T$ is the effective temperature of mode $z$ and $k_B$ is the Boltzmann constant. During the dynamics the modes are immersed in thermal environments with $T_1,T_2= 300$~K. This initial thermal distribution imbalance between $T_1$ and $T_2'$ minimises the thermal fluctuations $\langle p_{{z_2},0}^2 \rangle$, which otherwise obscure the noise-induced effect, while simultaneously minimising any unwanted back-action effects on the particle whose noise drives the noise-induced effect. While the technical details of this preparation depend strongly on the chosen platform, we suggest a proof-of-principle state preparation scheme in SM Note 4.

In Fig.~\ref{fig1p2}a we show the noise induced motion for this parametrically symmetric case (grey). The classical simulations are carried out with the parameters $\omega_i= 50$~kHz, $m_1=m_2=8\times 10^{-17}$~kg, $\kappa = 2.3\times 10^{-24}$~N~m${}^2$ and $d=3$~$\mu$m, inspired by optical levitation~\cite{Brzobohaty2023,Rieser2022,reisenbauer_non-hermitian_2024,deplano_coulomb_2024} and enriched with magnetic levitation~\cite{Hofer2023,Gutierrez2023} platforms in mind as they operate with a wider trapping frequency range. No displacement occurs in $z_1$, however the noise grows rapidly. The mean momentum of the second particle, however, experiences a positive sharp increase away from zero as well as an increase in noise. This displacement is not critically outperformed by noise, as seen in the signal-to-noise ratio ($\mathrm{SNR}$) in Fig.~\ref{fig0}b(i) (empty circles). This is a sign of a direct noise-induced coherent effect. Importantly, the $\mathrm{SNR}$ {\it grows} with increasing $\sigma_{z_{1,0}}$ (full circles) and can saturate the maximum of $1/\sqrt{2}$ at the cost of large noise in the initial position of $z_1$. 

This effect can be explained by examining the Langevin equations of motion for the third order Coulomb term. These are given by
\begin{widetext}
\begin{equation}\label{eq4}
    \begin{aligned}
        m_2 \ddot{z}_2(t)+m_2\Gamma \dot{z}_2(t)&\approx\frac{3\kappa}{ d^4}\left( z_1^2(t) + z_2^2(t)\right) - \left(m_2\omega_2^2 +\frac{6\kappa}{ d^4}z_1(t) \right)z_2(t) + \sqrt{2\Gamma k_B T_2} \xi_2(t),\\
        m_1\ddot{z}_1(t)+m_1\Gamma \dot{z}_1(t)&\approx - \frac{3\kappa}{ d^4} \left( z_2(t)^2 + z_1^2(t) \right) - \left(m_1\omega_1^2 -\frac{6\kappa}{d^4}z_2(t)\right)z_1(t) +\sqrt{2\Gamma k_B T_1} \xi_1(t)\,,
    \end{aligned}
\end{equation}
\end{widetext}
where $\Gamma$ is the drag coefficient whose value and phenomenological origin depend upon implementation and $\xi_1,\xi_2$ are independent zero-mean Gaussian white noises with $\langle \xi_i(t')\xi_i(t'') \rangle=\delta(t'-t'')$. 
In what follows we focus on the underdamped regime, with $\Gamma= 10^{-4}$~Hz, as  in the overdamped regime the nonlinear effect in momentum $p_{z_2}$ is not visible (see SM Note 1).
For parametric symmetry it is useful to discuss the dynamics using the mean value approximation, by reducing the two-body interaction into a one-body problem by virtue of the effective potentials $\tilde{V}(z_2)= - 3\kappa \langle z_1 \rangle ^2 z_2/d^4 + \tilde{\omega}_2 z_2^2 - \kappa z_2^3/d^4$ and $\tilde{V}(z_1)= 3\kappa \langle z_2\rangle^2z_1/d^4 + \tilde{\omega}_1 z_1^2 + \kappa z_1^3/d^4$. 
In this framework, the unwanted back-action of $z_2$ on $z_1$ is understood as the mean displacement $\langle z_2 \rangle$ which (i) generates a drift in $z_1$, as visible from the first term in $\tilde{V}(z_1)$, and (ii) modifies the frequency of the harmonic confinement of $z_1$ via $\tilde{\omega}_1 = m_1 \omega_1^2/2 - 3\kappa \langle z_2 \rangle /d^4$. 
The imbalance in the initial noise properties minimises both back-action contributions at short transients $t<20$~$\mu$s. That is, low temperature $T_2'$ makes the noise-induced shift generated by the cubic potential in $\tilde{V}(z_2)$ negligible, keeping the position below the critical value of $\langle z_2 \rangle \approx 1$~$\mu$m (calculated for the parameters used in numerical simulation) after which the effective frequency $\tilde{\omega}_1$ becomes negative and the harmonic confinement becomes an inverted quadratic potential, leading to unstable diverging trajectories.
However, the noise in $z_1$ is still substantially increasing in time, and can in general complicate both predictions and applications of nonlinearities.
For large initial noise, as visible in Fig.~\ref{fig1p2} (a, grey), the higher order nonlinear terms of the Coulomb interaction make for a positive back-action $z_2$ on $z_1$. It results in a decrease of the noise of $z_1$, even below that of its initial thermal state.

When the reciprocity in the nonlinear effect is further broken by either tuning the mass (at fixed frequency) or frequency (at fixed mass) of particle 1, the fluctuations in $z_1$ are modified and the properties of the coherent motion are altered.
The initial thermal state of $z_1$ is determined by its dynamical variables $m_1$, $\omega_1$, and local environmental temperature $T_1$ via the standard deviation $\sigma_{z_1,0} = \sqrt{k_B T/m_1\omega_1}$.  
Changing the mass or frequency to pursue the symmetry breaking techniques results in a modification of the initial fluctuations. We therefore fix the noise properties to $\sigma_{z_1,0}=30$~nm to observe the effect arising from the nonlinear interaction with constant initial noise across different parameters regimes. 
That is, $z_1$ is not prepared in an equilibrium state of its local oscillator but rather in an out-of-equilibrium thermally squeezed state.

Trapping a massive particle $m_1\gg m_2$ minimises its kinetic term $p^2_{z_1}/2m_1$, and as a result the position does not move away from its initial mean condition $\braket{z_1} \approx \braket{z_{1,0}}$. That is, the back-action is negligible for short transients, as visible. 
Differently than parametric symmetry, the additional imbalance of the mass reshapes the noise evolution of $z_1$, keeping it close to its initial distribution $\sigma_{z_{1,0}}$ for low initial fluctuations.
However, as visible in Fig.~\ref{fig1p2} (a, orange), large initial noise promotes a positive back-action loop over time, decreasing the fluctuations of $z_1$ below the initial thermal state. It is similar to the parametric symmetry (grey) back-action effect, however its effect on particle 2 showcases a larger displacement although with larger fluctuations (orange shaded area).
The resulting $\mathrm{SNR}$ in Fig.~\ref{fig0}b(i), while increasing even at low fluctuations (empty circles), only saturates the bound for larger initial noise $\sigma_{z_{1,0}}$ (full circle). 
The short transient of $p_2$ evolves as $p_{z_2}(t') \approx p_{z_2,0} + 3\kappa \int_0^{t'} ds' \, z_1^2(s')/d^4 - m_2\omega_2^2 z_{2,0} t'$ in the limit of zero damping $\Gamma=0$. 
For mass tuning it leads to the following moments
\begin{equation}\label{eq5}
    \langle p_2(t) \rangle \approx \frac{3 \kappa t \sigma_{z_1,0}^2}{ d^4}, \quad
    \mathrm{SNR} = \frac{\langle p_2 \rangle}{\sigma_{p_2}} \approx \frac{1}{\sqrt{2}}.
\end{equation}

Alternatively, trapping particle 1 in a stiffer harmonic potential $\omega_1 \gg \omega_2$ confines its noise dynamics to that of a harmonic oscillator for short transients. Its evolution is described by coherent oscillations approximately described by $z_1 \approx z_{1,0} \cos(\omega_1 t)$ under the assumption of vanishing initial velocity $\dot{z}_{1,0}=0$. As visible in Fig.~\ref{fig1p2} (a, blue) this oscillatory evolution dominates over the back-action for times of few tenths of microseconds.
This added imbalance of frequency negatively impacts the dynamics, generating a lower momentum drift $\langle p_{z_2} \rangle$, and a lower signal-to-noise ratio output Fig.~\ref{fig0} (b(i), blue) at small initial noise (empty circles), but it too saturates the $1/\sqrt{2}$ bound at large initial noise $\sigma_{z_{1,0}}$ (full circles).
For tuning frequencies, the moments of momentum $p_2$ approach
\begin{equation}\label{eq6}
    \langle p_{z_2}(t) \rangle \approx \frac{3\kappa}{4 d^4 \omega_1} \left[ 2\theta + \sin(2\theta)\right] \sigma_{z_1,0}^2,\quad
    \mathrm{SNR} \approx \frac{1}{\sqrt{2}},
\end{equation}
where $\theta = \omega_1 t$.

\begin{figure*}[t!]
\centering
\includegraphics[width=\textwidth]{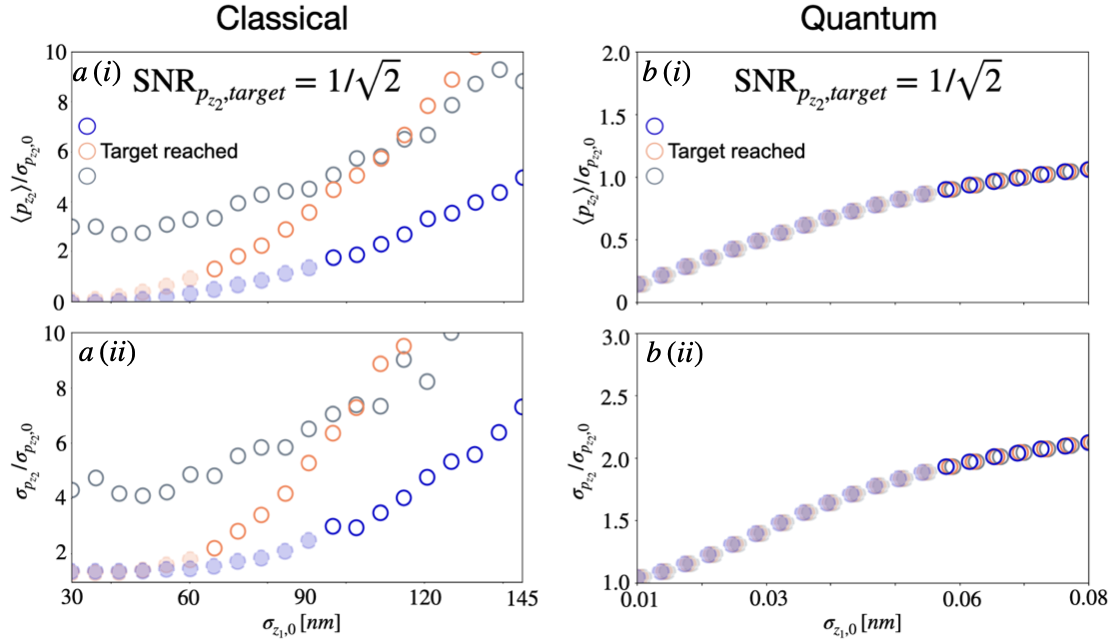}
\caption{\label{fig1}\small{{\bf Noise/Uncertainty induced momentum under noise confinement.} (a): The output displacement $\langle p_{z_2} \rangle$ (i), and standard deviation $\sigma_{p_{z_2}}$ (ii) at the target signal-to-noise ratio $\mathrm{SNR}_{p_{z_2}}=1/\sqrt{2}$ are plotted for the stochastic classical dynamics against the input noise $\sigma_{z_1,0}$. 
At low initial input noise the parametric symmetry (grey) always reaches the target. At large input noise all regimes reach the target, but breaking the symmetry via tuning frequency (blue) provides the least noise output, and thus even smaller momentum displacement.
Breaking symmetry by tuning mass (orange) is useful only between noise input $60 \lesssim \sigma_{z_1,0} \lesssim 100$~nm.
(b): The output displacement $\langle p_{z_2} \rangle$ (i), and standard deviation $\sigma_{p_{z_2}}$ (ii) at the target $\mathrm{SNR}$ for different initial uncertainty $\sigma_{z_1,0}$. All regimes reach the target, but at different times. For parametric symmetry (grey) and tuning  frequency (blue) the target is reached at $t\approx 1\, \mu$s, while for tuning mass (orange) the target is reached at larger times $t \approx 2\, \mu$s. Regardless of the required time, when the target is reached, all regimes produce the same displacement and standard deviation output making the parametric symmetry (grey) the preferred strategy to reach the target with the minimum noise cost. Note, breaking symmetry requires extra squeezing to reach the same initial noise input.
The dashed filled circles record the value of momentum displacement and standard deviation when the target $\mathrm{SNR}$ is not reached
.}}
\end{figure*}

In the low noise limit, the added symmetry breaking lowers the $\mathrm{SNR}$ relative to the symmetric case (see Fig.~\ref{fig0}b(i), empty circle).
To reach it, an extra cost of increasing initial noise $\sigma_{z_1,0}$ is required. 
Specifically, for tuning stiffness, the same $\mathrm{SNR}$ is reached at smaller displacement (see Fig.~\ref{fig1p2}a, blue), while for tuning mass the $\mathrm{SNR}$ saturation is obtained with a much larger displacement (orange), making it a favorable strategy.
This is further highlighted in Fig.~\ref{fig1}a, where a target $\mathrm{SNR}=1/\sqrt{2}$ is fixed, and the output momentum displacement $\langle p_{z_2} \rangle$ (top), and standard deviation $\sigma_{p_{z_2}}$ (bottom) are plotted against input noise cost $\sigma_{z_1,0}$. 
It shows that for initial noise below $\sigma_{z_1,0} \lesssim 100$~nm, the parametric symmetry (grey) harnesses the noise of $z_1$ through the Coulomb interaction more efficiently. 
However, for initial noise input beyond $\sigma_{z_1,0} \gtrsim 100$~nm, tuning frequency (blue) generates the same signal-to-noise ratio with smaller output noise and momentum drift, while tuning mass (orange) reaches it with larger momentum drift. Fig.~\ref{fig1} shows that the best strategy, for large initial noise, is to break the symmetry by tuning mass to reach the target signal-to-noise with the largest momentum drift.

\subsubsection{Quantum Uncertainty-Induced Momentum}

As we decrease the initial noise to the ground state extension, that is $\sigma_{z_{1,0}} = 0.01$~nm, the nonlinear effect in the stochastic framework described by Eq.\eqref{eq4} vanishes (see SM Note 2).
Operating in the quantum regime of Eq.~\eqref{eq4}, using pure initial states, an analog of the previous noise-induced phenomena comes directly from the quantum fluctuations in $z_1$. As expected, it is sufficient to produce momentum displacement on $z_2$, as visible in Fig.~\ref{fig1p2}b.

In this section, the numerical simulation are performed directly on Eq.~\eqref{eq4}, see SM Note 2, using the same parameters outlined in the previous section. All quantities are rescaled by the standard deviation of the initial state.
For parametric symmetry, the quantum fluctuations of $z_1$ induce a small drift in the momentum, further enhanced by the quantum fluctuation of $z_2$ through the cubic nonlinearity, reaching  $\langle p_{z_2} \rangle \approx 3\kappa (\sigma_{z_{1,0}}^2 + \sigma_{z_{2,0}}^2 )t/d^4$. The unwanted back-action force is too small to induce instability, therefore the noise of $z_1$ does not increase larger than its reference state. 
It drives the momentum $\langle p_{z_2} \rangle$ with an
increasing $\mathrm{SNR}$ (panel $c_1$, empty circles) that does not reach the maximum of $1/\sqrt{2}$, as for times $t>2\,\mu$s the cubic nonlinearity dominates both dynamics and the back-action strongly drives $z_1$ to instability. It enhances the noise of $p_2$ beyond the initial ground state, thus resulting in a drop of the $\mathrm{SNR}$.
The conservative symmetry breaking strategies, introduced for the stochastic dynamics, result in a qualitatively similar time evolution (see Fig.~\ref{fig1p2}b, blue and orange).

The initial standard deviation $\sigma_{z_1,0} =\sqrt{ \hbar/(2m_1\omega_1)}$ is calculated using the same  parameters of the stochastic system. For parametric symmetry, i.e., $m_1 = 8 \times 10^{-17}$~kg and $\omega_1 = 50$~kHz, the initial standard deviation results in $\sigma_{z_1,0}=0.01$~nm.
For tuning frequency, i.e., $\omega_1 = 2500$~kHz, and tuning mass $m_1 = 8\times 10^{-16}$~kg, the initial standard deviations assume different values. Respectively $\sigma_{z_1,0}=0.001$~nm, and $\sigma_{z_1,0}=0.003$~nm.
To observe only the effects of the nonlinear interaction given by the dynamics, under the same initial noise conditions, the initial state of $z_1$ is squeezed by a factor  $\xi= -\log(\sigma_{trg}\sqrt{m_1\omega_1/\hbar})/2$ to reach the unified target standard deviation of $\sigma_{trg} = 0.01$~nm. 
That is, the position variance is amplified $\sigma_{z_1,0} = \xi \sqrt{\hbar/(2m_1\omega_1)}$ by $\xi$, while the momentum variance is attenuated by the inverse amount $\sigma_{p_{z_1,0}} = \sqrt{\hbar m_1 \omega_1/2}/\xi$.

When the initial ground state of $z_1$ is further squeezed in momentum, it realises a larger drift reaching a $\mathrm{SNR}=1/\sqrt{2}$ at short transients (Fig.~\ref{fig0}, panel b$_2$).
Notice that parametric symmetry (grey) and tuning frequency (blue) are subjected to instability for times larger than $t\approx 1\, \mu$s, resulting in a drop of the signal-to-noise ratio, while tuning mass (orange) is not yet affected by it.

For tuning mass $m_1 \gg m_2$ the short transient of the moments of $p_2$ approach
\begin{equation}\label{stats_px}
    \begin{aligned}
        \langle p_{z_2} \rangle &\approx \frac{3\kappa t \sigma_{z_1,0}^2}{d^4},\\
        \mathrm{SNR}_{p_{z_2}} &\approx \frac{1}{\sqrt{2}}\Big[ 1 + \frac{m_2^2 \omega_2^4 d^8 \sigma_{z_2,0}^2}{18 \kappa^2 \sigma_{z_1,0}^4} + \frac{d^8 \sigma_{p_{z_2},0}^2}{18\kappa^2 \sigma_{z_1,0}^4 t^2} \Big]^{-\frac12}.
    \end{aligned}
\end{equation}

For short transients, the ground state momentum noise $\sigma_{p_{z_2},0}^2$ prevents the signal-to-noise ratio from reaching the $1/\sqrt{2}$ bound. At larger time it becomes negligible, leaving the position noise $\sigma_{z_2,0}^2$ as the dominant limiting term of the evolution. For fixed $m_2,\omega_2$, an amplification of position noise $\sigma_{z_1,0}^2 = \xi \sigma_{z_1,0}^2$ by $\xi$ allows to reach a $\mathrm{SNR}=1\sqrt{2}$ as visible in Fig.~\ref{fig0}b(ii) , full orange circle. That is, the initial state is further squeezed in momentum.
Squeezing the position noise $\sigma_{z_2,0}^2$ can in principle improve upon the signal-to-noise ratio of Eq.~\eqref{stats_px}. However, the complementary amplification of momentum noise $\sigma_{p_{z_2},0}^2$ increases the back-action to particle $1$ at larger times, thus leading to the divergence quicker.

For tuning frequency $\omega_1 \gg \omega_2$, the noise of $z_1$ is confined in a stiffer harmonic bound, and its dynamics is described as $z_1 \approx z_{1,0} \cos(\theta)$, with $\theta=\omega_1 t$.
In this regime the dynamics evolves similarly to that of the parametric symmetry as visible in Fig.~\ref{fig0}(ii), blue and grey circles. Its momenta evolve in short transients according to
\begin{equation}\label{eq7}
    \begin{aligned}
        \langle p_{z_2} \rangle &\approx \frac{3\kappa \sigma_{z_1,0}^2}{4 d^4 \omega_1} \left[ 2\theta + \sin(2\theta) \right],\\
        \mathrm{SNR}_{p_{z_2}} &\approx \frac{1}{\sqrt{2}} \Big[1+ \frac{\left(8 \omega_2^2 d^8 \sigma_{p_{z_2},0}^2 + 8 \omega_1^2 d^8 t^2 \sigma_{z_2,0}^2 \right)}{9\kappa^2 \sigma_{z_1,0}^4 \left[2\theta + \sin(2\theta) \right]^2} \Big]^{-\frac12}.
    \end{aligned}
\end{equation}

For an initial ground state $\sigma_{z_1,0}^2 = \hbar/(2m_1\omega_1)$, the momentum and position noise of the initial state of particle $2$, i.e., $\sigma_{z_2,0}^2,\sigma_{p_{z_2},0}^2$ hinders the uncertainty-induced effect from reaching the $\mathrm{SNR}=1/\sqrt{2}$ bound at short transients. For longer time the instability from the cubic potential, not accounted for in Eq.~\eqref{eq7}, dominates the dynamics resulting in a drop of the signal-to-noise ratio.
Similar to tuning mass, to reach the signal-to-noise bound for short transients, the amplification of position noise $\sigma_{z_1,0}^2$ is required, as visible in Fig.~\ref{fig0}b(ii), full blue circles. Moreover, squeezing the position noise $\sigma_{z_2,0}$ results in faster divergence, similar to the case of tuning mass. 

The hidden cost of parametric symmetry breaking lies in the preparation of the initial state, which requires squeezing of the ground state. 
That is, to have comparable noise in position of $\sigma_{z_1,0}=0.01$~nm after tuning mass and frequency the ground state must be squeezed, by $\xi=1.15$ and $\xi = 1.96$ respectively.
For the large noise case, a further amplification by freefall is then used to reach the position noise in all regimes of $\sigma_{z_1,0}=0.08$~nm.
In Fig.~\ref{fig1}b, the target $\mathrm{SNR}=1/\sqrt{2}$ is reached equally by all regimes for initial input noise $\sigma_{z_1,0}\gtrsim 0.06$~nm, but at different times. 
Ultimately, the parametric symmetry emerges as the best regime, as it does not require any extra costs, i.e., the initial squeezing to $\sigma_{z_1,0}=0.01$~nm, as is the case of the mass and frequency tuning regimes.

\section{Conclusion}

Experiments involving the interaction of trapped charged particles typically operate in the harmonic approximation~\cite{deng_quantum_2008,rudolph_force-gradient_2022,penny_sympathetic_2023}, where the motion decouples into oscillations along each coordinate axis. This can only result in Gaussian effects: single particle and multiparticle squeezing, and potentially, entanglement in the quantum regime. Here we have shown a proof-of-principle method to go beyond this approximation, showing that the inherently nonlinear effect of noise-induced coherent motion can be seen and explained using the first nontrivial nonlinear term in an expansion of the reciprocal Coulomb interaction between two charged particles.
This is observed in Fig.~\ref{fig0}b(i) by the non-reciprocal effect: improvement in $\mathrm{SNR}$ of one particle due to increased position thermal noise in another nonlinearly coupled particle. This effect persists into the quantum regime, where now the quantum uncertainty rather than classical noise drive the dynamics of the test particle to the same $\mathrm{SNR}$ bound of classical dynamics, at smaller momentum displacement. It is visible in Fig.~\ref{fig0}b(ii).

This method provides the first step to analysing one of the most basic two-particle nonlinear effects, as visible in Fig.~\ref{fig1}. Noise control is an important basic tool for
control of technologies that are inherently stochastic. Advantageously, this can be done autonomously through naturally occurring forces, lowering the engineering requirements to make such effects visible across a broad range of parameters and platforms.
It should be noted that nano-particles possess a natural advantage compared to ions as their mass-charge ratio favours observation of the short time nonlinear effects. That is, their high charge and high mass range capabilities allow for a stronger interaction tunability and slower divergence respectively.

To date, the most in-depth studies of motional nonlinear systems are of single-mode nonlinearities, possibly linked to other linear systems. Increasingly detailed studies of experimentally accessible two-particle nonlinear interactions will likely unveil many unexpected effects. Breaching further into genuinely multipartite systems combines the complexity of the many possible configurations (chains, triangles, lattices, or clusters)~\cite{Lu2023,Frattini2017,Bykov2022} with that of a nonlinear interaction distributed across multiple particles, likely hiding exciting nonlinear effects. This then opens the possibility to further exploit such nonlinearity in naturally occurring interactions. 

\section{Acknowledgements}
We acknowledge the project GA23-06224S of the Czech Science Foundation, EU and MEYS Czech Republic No. CZ.02.01.01/00/22\_008/0004649 (QUEENTEC). R.F.
also acknowledges funding from the MEYS of the Czech Republic (Grant Agreement 8C22001). Project SPARQL has received funding from the European Union’s Horizon 2020 Research and Innovation Programme under Grant Agreement no. 731473 and 101017733 (QuantERA).

\section{Data and Code Availability}
Data sharing not applicable to this article as no datasets were generated or analysed during the current study. Code available upon reasonable request.

\section{Competing Interests}
The authors declare no competing interests.

\section{Author Contributions}
LO performed the numerical simulations and analytical solutions to the classical dynamics and DM performed the quantum mechanical calculations. Both received theory inputs from RF. RF conceived and supervised the project. All authors contributed to the analysis of the results and composition of the article.

\bibliography{references.bib}

\end{document}